\documentclass[fleqn, usenatbib]{mnras}

\usepackage{newtxtext}

\usepackage[T1]{fontenc}
\usepackage{ae,aecompl}

\usepackage{graphicx}	
\usepackage{amsmath}	
\usepackage{amssymb}	
\usepackage{subfig}
\usepackage[font=footnotesize]{caption}
\usepackage{graphicx}
\usepackage{wrapfig}
\usepackage{gensymb}
\usepackage{comment}
\usepackage{multirow}
\usepackage{tabularx}
\usepackage{footnote}
\usepackage[dvipsnames]{xcolor}
\usepackage{stackengine}
\usepackage{multirow}
\usepackage{tablefootnote}
\usepackage{enumitem}
\usepackage{longtable}
\usepackage{threeparttablex}
\usepackage{hhline}
\usepackage{bm}		
\usepackage{pdflscape}	

\def\maxififth{MAXI~J1535--571}
\def\maxithirt{MAXI~J1348--630}
\def\maxieight{MAXI~J1820$+$070}

\def\hh{H1743--322}
\def\gx{GX~339--4}

\newcommand{\tu}{\textup}

\title[Radio/X-ray correlation in \maxithirt{}]{The hybrid radio/X-ray correlation of the black hole transient \maxithirt{}}

\author[F. Carotenuto et al.]{F. Carotenuto,$^{1}$\thanks{E-mail: francesco.carotenuto@cea.fr}
S. Corbel,$^{1,2}$
E. Tremou,$^{3}$
T. D. Russell, $^{4}$
A. Tzioumis,$^{5}$
R. P. Fender,$^{6,7}$
\newauthor
P. A. Woudt,$^{7}$
S. E. Motta,$^{6}$
J. C. A. Miller-Jones,$^{8}$
A. J. Tetarenko,$^{9}$
G. R. Sivakoff$^{10}$
\\
$^{1}$AIM, CEA, CNRS, Universit\'{e} de Paris, Universit\'{e} Paris-Saclay, F-91191 Gif-sur-Yvette, France\\
$^{2}$Station de Radioastronomie de Nan\c cay, Observatoire de Paris, PSL Research University, CNRS, Univ. Orl\'eans, 18330 Nan\c cay, France\\
$^{3}$LESIA, Observatoire de Paris, CNRS, PSL Research University, Sorbonne Universit\'{e}, Universit\'{e} de Paris, Meudon, France\\
$^{4}$INAF, Istituto di Astrofisica Spaziale e Fisica Cosmica, Via U. La Malfa 153, I-90146 Palermo, Italy\\
$^{5}$Australia Telescope National Facility, CSIRO, PO Box 76, Epping, New South Wales 1710, Australia\\
$^{6}$Astrophysics, Department of Physics, University of Oxford, Keble Road, Oxford OX1 3RH, UK\\
$^{7}$Inter-University Institute for Data-Intensive Astronomy, Department of Astronomy, University of Cape Town, Private Bag X3, Rondebosch  7701, South Africa\\
$^{8}$International Centre for Radio Astronomy Research, Curtin University, GPO Box U1987, Perth, WA 6845, Australia\\
$^{9}$East Asian Observatory, 660 N. A'\!oh$\bar{o}$k$\bar{u}$ Place, University Park, Hilo, Hawaii 96720, USA\\
$^{10}$Department of Physics, University of Alberta, CCIS 4-181, Edmonton, AB T6G 2E1, Canada\\
}
\date{Accepted XXX. Received YYY; in original form ZZZ}

\pubyear{2021}

\begin{document}
\label{firstpage}
\pagerange{\pageref{firstpage}--\pageref{lastpage}}
\maketitle

\begin{abstract}
\noindent
Black hole low mass X-ray binaries in their hard spectral state are found to display two different correlations between the radio emission from the compact jets and the X-ray emission from the inner accretion flow. Here, we present a large data set of quasi-simultaneous radio and X-ray observations of the recently discovered accreting black hole \maxithirt{} during its 2019/2020 outburst. Our results span almost six orders of magnitude in X-ray luminosity, allowing us to probe the accretion-ejection coupling from the brightest to the faintest phases of the outburst. We find that \maxithirt{} belongs to the growing population of \textit{outliers} at the highest observed luminosities. Interestingly, \maxithirt{} deviates from the outlier track at $L_{\tu{X}} \lesssim 7 \times 10^{35} (D/2.2 \ \tu{kpc})^2$ erg s$^{-1}$ and ultimately rejoins the standard track at $L_{\tu{X}} \simeq 10^{33} (D/2.2 \ \tu{kpc})^2$ erg s$^{-1}$, displaying a \textit{hybrid} radio/X-ray correlation, observed only in a handful of sources. However, for \maxithirt{} these transitions happen at luminosities much lower than what observed for similar sources (at least an order of magnitude). We discuss the behaviour of \maxithirt{} in light of the currently proposed scenarios and we highlight the importance of future deep monitorings of hybrid correlation sources, especially close to the transitions and in the low luminosity regime.

\end{abstract}


\section{Introduction}
\label{sec:Introduction}
Black holes (BH) low mass X-ray binaries (LMXBs) are binary systems comprising a BH that accretes matter from a low mass companion star. Occasionally, these systems enter outburst phases during which the X-ray luminosity is greatly variable and several transitions between distinct accretion states are observed (e.g. \citealt{Remillard_xrb, Corral_santana, Tetarenko_2016, Belloni_Motta_2016}) A strong coupling exists between accretion and ejection in BH LMXBs, as the relativistic jets powered by these sources are highly dependent on the state of the accretion disk. From an observational point of view, BH LMXBs in the hard spectral state are characterised by a non-linear correlation between the radio and X-ray luminosities ($L_{\textup{R}} \propto L_{\textup{X}}^{\beta}$, e.g. \citealt{Hannikainen_1998, Corbel_2000, Gallo_2003}).
Based on scale-invariant properties of black hole accretion and jet production, this correlation has also been extended to AGNs by including the mass as an additional parameter \citep{Merloni_2003, Falcke_2004, Plotkin_2012}. BH LMXBs are divided in two populations which lie on two different tracks of the radio/X-ray diagram. 
Originally, the few "historical" BH sources were found to lie on a track labeled as \emph{standard}, with a power law index $\beta \sim$ 0.6 (such as \gx, V404 Cyg and now also \maxieight{}, \citealt{Corbel_2000, Corbel_2003, Corbel2013_corr, Gallo_2003, Bright, Shaw_2021}). These sources appear to maintain this correlation down to the quiescent level \citep{Corbel2013_corr, Gallo2019, Tremou2020}. As new observations accumulate, more and more sources (labeled \emph{outliers}) were found to lie on a different branch, well below the standard track, with a steeper slope $\beta \gtrsim 1$ \citep{Coriat}. We note that in some works the two populations are referred to as \textit{radio-loud} and \textit{radio-quiet}, one of the two ways to interpret the data, see \cite{Coriat} for more details. Outliers could be characterized by more negative radio spectral indices \citep{Espinasse} and lower rms variability \citep{Dincer, Motta_2018} with respect to standard track sources, which may be indications for understanding the nature of the two different tracks. While the reasons for the the existence of this dichotomy are still unclear, several explanations for the observed scenario have been proposed. A physical difference in the disk-jet coupling may be responsible for the two groups. This could originate from the structure of the inner accretion flow or alternatively the properties of jets might differ between the two tracks, causing different levels of radio emission  (\citealt{Coriat}; see also Section \ref{Discussion} for more details).
The existence of the two tracks has also been questioned statistically 
(e.g. \citealt{Gallo_2014, Gallo_2018}) if one considers the whole sample of sources. However, it is particularly important to precisely track the behaviour of specific sources, such as \hh{} \citep{Coriat, Williams_2020}, for which the path on the diagram is very clear, to pinpoint the overall behaviour which may be masked in the whole sample due to overlap of sources and possible transition between the two tracks.

In this Letter we present radio and X-ray observations of \maxithirt{} to highlight its behaviour on the radio/X-ray diagram. Thanks to the deep and long term coverage of our multi-wavelength monitoring, we are able to probe the source behaviour on the radio/X-ray diagram in great detail, from the bright phases, down to the lowest level of emission in outburst, when the source is close to its quiescent state. So far this has been possible for only a limited number of sources, especially over a single outburst (see for instance~\citealt{Corbel2013_corr} and~\citealt{Shaw_2021}). \maxithirt{} is a new BH LMXB discovered in January 2019 \citep{Yatabe2019, Tominaga_1348}, located at a distance $D = 2.2^{+0.6}_{-0.5}$ kpc \citep{Chauhan2021}. We note that \cite{Lamer_2021} recently reported a distance $D = 3.39 \pm 0.34$ kpc from observations with eROSITA. While in this paper we assume $D = 2.2$ kpc, to include the second distance measurement we quote all derived luminosities with a factor $(D/2.2 \ \tu{kpc})^2$. In \cite{Carotenuto2021} we have presented the full X-ray and radio monitoring of the source during its 2019/2020 outburst. The source first completed a whole cycle in the Hardness Intensity Diagram (HID), and then exhibited a sequence of hard state re-brightenings (e.g. \citealt{DRussell_atel_opt_2, Yazeedi_atel_2, Pirbhoy_atel}). Our radio observations detected and monitored the rise, quenching and re-activation of compact jets through the different phases of the outburst. In addition, single-sided and resolved discrete ejecta were detected \citep{Carotenuto2021}. Two jet components were launched $\sim$2 months apart and both displayed a very high proper motion ($\gtrsim 100$ mas day$^{-1}$).

\vspace{-4.em}

\section{Observations}
\label{sec:Observations}

\maxithirt{} has been monitored with MeerKAT \citep{Jonas2016, Camilo2018} at 1.28 GHz, as part of the ThunderKAT Large Survey Programme \citep{ThunderKAT}, and with the Australia Telescope Compact Array (ATCA) at 5.5 and 9 GHz. On the X-ray side, \maxithirt{} was regularly monitored by the Neil Gehrels \emph{Swift} Observatory \citep{Gehrels} with the X-ray Telescope (XRT, \citealt{Burrows_xrt}), and by the Monitor of All-sky  X-ray  Image (MAXI, \citealt{Matsuoka_maxi}). The full observing campaign on \maxithirt{} has already been presented in \cite{Carotenuto2021} and we refer to that work for details on the radio and X-ray data processing. 
In this work we also include two additional detections of \maxithirt{} during its seventh hard state re-flare (September-October 2020, \citealt{MAXI_Atel_SepOct2020, Carotenuto_SepOct2020_Atel}), from observations performed with MeerKAT, ATCA and \textit{Swift}/XRT and not reported in \cite{Carotenuto2021}.

\section{The Radio/X-ray correlation of \maxithirt{}}
\label{sec:The Radio/X-ray correlation of MAXIJ1348}

To characterise the behaviour of \maxithirt{} in the radio/X-ray diagram, we selected epochs with quasi-simultaneous ($\Delta t \leq 24$ h) radio and X-ray observations between MJD 58509 and 58522, and after MJD 58597, during which the system was in the hard and intermediate state (see \citealt{Zhang2020} and \citealt{Carotenuto2021} for details on the state transitions). For each epoch we converted the measured radio flux density $S_{\nu}$ to the 5 GHz monochromatic luminosity $L_{\tu{R}} = 4\pi D^2  \nu S_{\nu}$, assuming a distance $D = 2.2$ kpc and using either the measured spectral index $\alpha$ (see Table \ref{tab:sample_x_radio_corr}) when available, or the average spectral index $\langle \alpha \rangle = 0.14 \pm 0.01$. The unabsorbed 1--10 keV X-ray flux $S_{\tu{X}}$ obtained from \emph{Swift}/XRT was converted to the integrated luminosity $L_{\tu{X}} = 4\pi D^2  S_{\tu{X}}$. For simultaneity, we selected \emph{Swift} epochs taken less than 24 hours before or after the corresponding radio observations. When this was not possible, we interpolated the X-ray flux from a exponential fit to the smoothly-evolving \emph{Swift}/XRT light curve, adding a conservative 10\% error on the interpolated X-ray fluxes and checking the consistency with the simultaneous MAXI data. This leads to a sample of 44 measurements on the radio/X-ray diagram, with a total of 39 detections of \maxithirt{} in radio and X-rays.

\begin{figure*}
\begin{center}
\includegraphics[width=0.95\textwidth]{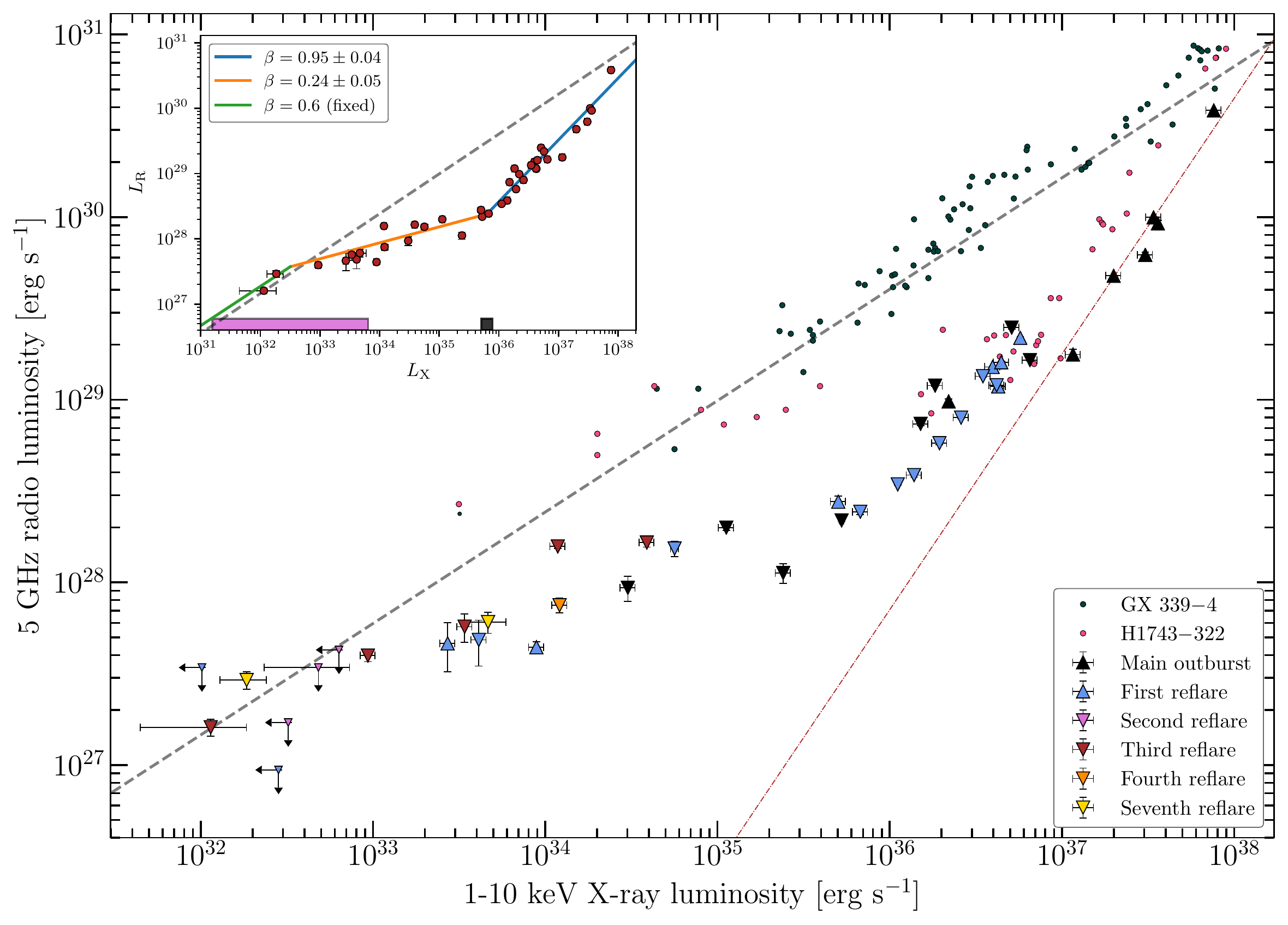}
\caption{\maxithirt{} radio/X-ray diagram during its 2019/2020 outburst. We assume $D = 2.2$ kpc \citep{Chauhan2021}. Points from different phases of the outburst are plotted with different colors, and triangles pointing upwards or downwards differentiate between, respectively, rise and decay. The red and grey dashed lines represent, respectively, the $\beta = 1.4$ and $\beta = 0.6$ correlations (see Section \ref{sec:Introduction}), and are shown for illustrative purposes. The source transitions between tracks while going from higher to lower luminosities. For comparison, \hh{} \citep{Coriat} and \gx{} \citep{Corbel2013_corr, Tremou2020} are shown on the diagram with smaller size points. A tentative fit to the same data is shown in the inset in the top-left corner. The magenta and grey rectangles represent the lower and upper bounds on the \textit{x}-axis of, respectively, the $L_{\tu{stand}}$ and $L_{\tu{trans}}$ confidence intervals (see Section \ref{sec:The Radio/X-ray correlation of MAXIJ1348}).}
\label{fig:x_radio_corr}
\end{center}
\end{figure*}

A sample of the flux values is reported in Table \ref{tab:sample_x_radio_corr}, while the full data set is provided as online supplementary material, and the radio/X-ray measurements of \maxithirt{} are shown in Figure \ref{fig:x_radio_corr}. We note that our monitoring allowed us to probe the behaviour of \maxithirt{} over six orders of magnitude in the X-ray luminosity and almost four in its radio luminosity, which so far has rarely been possible for XRBs. Moreover, our measurements include three detections at $L_{\textup{X}} < 10^{33} (D/2.2 \ \tu{kpc})^2$ erg s$^{-1}$, which are of key importance to probe the accretion/ejection connection at the lowest luminosities. To allow a direct comparison, measurements of \gx{} \citep{Corbel2013_corr, Tremou2020} and \hh{} \citep{Coriat} are also shown, as the two sources are representative of the standard and outlier branches, respectively.
\maxithirt{} immediately appeared to be incompatible with the standard track, being significantly fainter in radio than \gx{} below $L_{\textup{X}} < 10^{38} (D/2.2 \ \tu{kpc})^2$ erg s$^{-1}$. For $L_{\textup{X}}$ between $\sim$5 $\times 10^{35}$ and $10^{38} (D/2.2 \ \tu{kpc})^2$ erg s$^{-1}$ the source appeared to follow a steep correlation, globally consistent with the population of outliers. We note that the track followed during the decay of the main outburst was brighter in radio with respect to the rising phase, possibly similar to the rise and decay phases of the 2010/2011 outburst of \gx{} \citep{Corbel2013_corr}. For $L_{\textup{X}} \lesssim 5 \times 10^{35} (D/2.2 \ \tu{kpc})^2$ erg s$^{-1}$, \maxithirt{} transitioned to a flatter correlation. Unlike the main outburst, the rise and decay of the first reflare are consistent with the same track. On MJD 58614, 58621 and 58628 the radio core detected with MeerKAT was not completely separated due to the presence of discrete ejecta, hence we put a \emph{caveat} on the radio fluxes obtained from source fitting in those three epochs (the two detections with the lowest radio emission during the decay of the main outburst and the faintest radio detection during the rise of the first reflare, see Figure \ref{fig:x_radio_corr}. As $L_{\textup{X}}$ decreases, the source ultimately re-joins the standard track at $L_{\textup{X}} \simeq 10^{33} (D/2.2 \ \tu{kpc})^2$ erg s$^{-1}$ and is clearly detected down to $\sim$10$^{32} (D/2.2 \ \tu{kpc})^2$ erg s$^{-1}$, displaying a behaviour similar to \hh{} \citep{Coriat}, but transitioning between the two tracks at vastly lower radio and X-ray luminosities.

We fit our radio and X-ray data of \maxithirt{} using {\tt curve$\_$fit} from the SciPy package to obtain a tentative estimation of the correlation slopes. Visual inspection shows that, given the evolution of \maxithirt{} on the diagram, a single power law does not appear to be adequate to fit the whole data set. Therefore, we adopt a double-broken power law, making the distinction between a \textit{steep} branch followed when  $L_{\textup{X}} \geq L_{\textup{trans}}$, a \textit{flatter} correlation for $L_{\tu{stand}}  < L_{\textup{X}} < L_{\textup{trans}}$, and assuming that the source follows the standard track with $\beta = 0.6$ for $L_{\textup{X}} \leq L_{\textup{stand}}$. We note that this fit is for illustrative purposes, and the results should only be taken as indicative, given the significant dispersion of the data obtained by merging the various phases of the outburst. The results of the fit are shown in the inset of Figure \ref{fig:x_radio_corr}. For the high luminosity part, we obtain a slope $\beta_{\tu{steep}} = 0.95 \pm 0.04$ with $L_{\textup{trans}} = 6.3^{+1.5}_{-1.2} \times 10^{35} (D/2.2 \ \tu{kpc})^2$ erg s$^{-1}$, while the fit yields $\beta_{\tu{flat}} = 0.24 \pm 0.05$ for the flat branch. Since we fixed $\beta_{\tu{stand}} = 0.6$, the only parameter left is $L_{\textup{stand}}$, namely the transition luminosity between the flat branch and the standard track, which is not constrained by the fit, as we obtain: $L_{\textup{stand}} = 3.2^{+61.3}_{-3.0} \times 10^{32} (D/2.2 \ \tu{kpc})^2$ erg s$^{-1}$. While the other parameters of the fit depend weakly on the unconstrained value of $L_{\textup{stand}}$, this means that we cannot in principle rule out that the flat branch extends down to quiescence. However, we note that for \hh{} and Swift J1753.5--0127, the two sources more similar to \maxithirt{}, there is clear evidence that the standard track is followed in the low-flux regime \citep{Coriat, Williams_2020, Plotkin_1753}. Moreover, we note that points with $L_{\textup{X}} \gtrsim 10^{37} (D/2.2 \ \tu{kpc})^2$ erg s$^{-1}$ seem to be more consistent with the steeper correlation $\beta = 1.4$, which is also shown for illustrative purpose on Figure \ref{fig:x_radio_corr}.

\renewcommand{\arraystretch}{0.95}
\setlength{\tabcolsep}{8pt}
\setlength{\extrarowheight}{.05em}
\begin{table*}
\caption{Sample of the radio flux densities and unabsorbed X-ray fluxes selected for the radio/X-ray correlation presented in this work, along with the radio spectral index (when available) and the telescopes used for obtaining the data (see \citealt{Carotenuto2021}). Here we only show the rising phase of the main outburst, while the full table is available online as supplementary material. The radio spectral index is only computed for ATCA multi-frequency observations.}
\label{tab:sample_x_radio_corr}
\begin{tabular}{*{6}{c}}
\hhline{======}
MJD & Outburst phase & X-ray flux$^{\rm b}$ & Radio flux density & Spectral index & Telescopes\\
 & & 1--10 keV & $[$mJy$]$& $\alpha$ &\\
\hline

58509.9 & Main outburst $-$ rise    &  38.1  $\pm$  0.3   &   3.4   $\pm$ 0.1     & 0.02 $\pm$ 0.09            &     ATCA/\emph{Swift}\\     
58511.0 &           $-$             &  201   $\pm$  20    &   6.2   $\pm$ 0.4     & 0.1 $\pm$ 0.2              &     ATCA/\emph{Swift}$^{\rm a}$\\
58512.0 &           $-$             &  346   $\pm$  35    &   13.70  $\pm$ 0.05    &                            &     MeerKAT/\emph{Swift}$^{\rm a}$\\
58514.0 &           $-$             &  528   $\pm$  53    &   21.9  $\pm$ 0.8     & 0.18 $\pm$ 0.02            &     ATCA/\emph{Swift}$^{\rm a}$\\
58515.2 &           $-$             &  590   $\pm$  59    &   28.6  $\pm$ 0.1     &                            &     MeerKAT/\emph{Swift}$^{\rm a}$\\
58515.9 &           $-$             &  625   $\pm$  3     &   34.4  $\pm$ 1.4     &                            &     ATCA/\emph{Swift}\\
58519.9 &           $-$             &  1320  $\pm$  132   &   135   $\pm$ 1       & 0.155 $\pm$ 0.003          &     ATCA/\emph{Swift}$^{\rm a}$\\
\hline
\multicolumn{6}{l}{$^{\rm a}$ Interpolated X-ray flux.}\\
\multicolumn{6}{l}{$^{\rm b}$ In units of 10$^{-10}$ erg cm$^{-2}$ s$^{-1}$.}\\
\end{tabular}
\end{table*}

\vspace{-1.em}

\section{Discussion}
\label{Discussion}

We have presented a detailed view of the radio/X-ray correlation of \maxithirt{} during its 2019/2020 outburst. The source clearly belongs to the growing population of \emph{outlier} BHs. We traced the evolution of \maxithirt{} down to very low luminosities and observed it as it rejoined the standard correlation. Spanning six orders of magnitude in X-ray luminosity, the observations collected on \maxithirt{} constitute the most complete data set of an outlier obtained over a single outburst. 

\subsection{Evidence for an hybrid correlation}
\label{sec:Evidence for an hybrid correlation}

We find that \maxithirt{} is a new member of a restricted group of sources which appear to transition between the two known tracks. This particular behaviour, labeled \textit{hybrid} \citep{Xie_2016}, has already been observed (usually only partially) in \hh{} \citep{Coriat, Williams_2020}, Swift J1753.5--0127 \citep{Plotkin_1753}, XTE~J1752--223 \citep{Ratti_2012}, MAXI~J1659--152 \citep{Jonker_MAXI}, GRS~1739--278 \citep{1739_2020}, \maxififth{} \citep{Russell_1535, Parikh} and MAXI~J1631--472 \citep{Monageng_2021}. However, \maxithirt{} is, after \hh{}, only the second source in this regime to have such detailed monitoring, allowing the radio/X-ray behaviour to be so well constrained. Other sources classified as outliers have not been observed at such low luminosity (e.g.\ \citealt{Corbel_2004, Brocksopp_2005}). In fact, all outliers could also belong to the hybrid class (\citealt{Motta_2018}, see \citealt{Bahramian} for a global display of the diagram). The path of \maxithirt{} in the radio/X-ray diagram agrees well with \hh{} \citep{Jonker_2010, Coriat}. We note that, while both sources display a steep correlation at high luminosities and re-join the standard track before approaching quiescence, \maxithirt{} overall spans a broader range in $L_{\textup{R}}$ and $L_{\textup{X}}$, and is significantly fainter in radio than \hh{} in the flat part of the correlation. 
Moreover, the transition luminosity between the two correlations is harder to identify in \maxithirt{} than in \hh{}, for which \cite{Coriat} find $L_{\textup{trans}} \simeq 5 \times 10^{36}$ erg s$^{-1}$ assuming a distance of 8 kpc, corresponding to $\sim 5 \times 10^{-3} L_{\tu{Edd}}$, and for which the standard track is reached at $\sim 5 \times 10^{-5} L_{\tu{Edd}}$. This is higher than what is inferred for \maxithirt{} in this work. If we assume a 7 $M_{\odot}$ black hole \citep{Tominaga_1348, Carotenuto2021}, the indicative value of $L_{\textup{trans}} \simeq 7 \times 10^{35} (D/2.2 \ \tu{kpc})^2$ erg s$^{-1}$ corresponds to a transition happening at $\sim$7 $\times$ 10$^{-4} (D/2.2 \ \tu{kpc})^2 L_{\tu{Edd}}$, while \maxithirt{} would re-join the standard track at the low value of $\sim$10$^{-6} (D/2.2 \ \tu{kpc})^2 L_{\tu{Edd}}$. It is important to note that our monitoring was conducted during a single outburst (including the following reflares), while the data on \hh{} were accumulated over multiple outbursts \citep{Coriat, Williams_2020}.

\subsection{The radio-quiet hypothesis}
\label{sec:radio_quiet}

There is not yet a universal consensus on the origin of the two tracks and on what drives the change in correlation of a given source on the radio/X-ray diagram. As outlined by \cite{Coriat}, the two groups might be characterised by different levels of radio emission, possibly arising from different jet properties. Standard track sources and outliers could be labeled as, respectively, \textit{radio-loud} and \textit{radio-quiet}. As suggested by \cite{Motta_2018}, the existence of two populations on the diagram may be the result of geometric effects, rather than intrinsic differences between sources. Low inclination sources could appear radio-louder due to geometry and Doppler boosting, and thus would lie on the standard track. While we still lack a precise estimation of the \maxithirt{} inclination angle $\theta$, considerations on the discrete ejecta proper motion yield $\theta \lesssim 45\degree$ \citep{Carotenuto2021}. Hence, as a medium-low inclination source, geometric effects might not be suited to explain the fainter level of radio emission from \maxithirt{}, similarly to what has been found for \maxififth{} \citep{Russell_1535, Parikh}. However, misaligned jets are possible (e.g. \citealt{Miller-Jones2019}). 

Alternatively, physical differences between compact jets from different systems might be at the basis of the observed dichotomy on the diagram. Hints of a radio spectral difference between standard sources and outliers have been found by \cite{Espinasse}, possibly supporting this hypothesis. Compact jets powered by \maxithirt{} showed an average spectral index $\langle \alpha \rangle = 0.14 \pm 0.01$, which is more in agreement with a standard-track source than with an outlier. We tested the agreement with a Kolmogorov-Smirnov test on our spectral index distribution against the Gaussian distributions from \cite{Espinasse}, similarly to \cite{Williams_2020}. The null hypothesis was our distribution being consistent either with the standard or with the outlier distribution. For the outlier case, we find a p-value of $\sim$10$^{-5}$, while we obtain a p-value of 0.33 when we test our distribution against the standard one. Therefore, we are able to reject the null hypothesis in the first case, concluding that the jet spectral index distribution of \maxithirt{} is inconsistent with what is observed for the population of outliers. We cannot draw conclusions regarding the agreement with the distribution observed in standard sources.

If a physical difference in the jets can explain the two tracks, this could possibly imply a change in the properties or in the morphology of the jets when sources belonging to the hybrid correlation transition between the two groups, as argued in \cite{Koljonen}. This should produce observable effects, including, according to \cite{Espinasse}, a significant change in the radio spectral index distribution between the tracks. For \maxithirt{} we could not constrain the spectral index at low luminosities, hence we cannot confirm or rule out a potential radio spectral difference between the two tracks, but such effect is not observed in other sources (e.g. \citealt{Shaposhnikov_2007}), questioning the validity of this scenario. 
Alternatively, outliers could be characterised by compact jets with higher magnetic fields with respect to standard track sources \citep{Peer_Casella_2009}. While we have no measurement on the jet's magnetic field in \maxithirt{}, an evolution of the magnetic field throughout the outburst, possibly increasing with the accretion rate and, thus, with $L_{\tu{X}}$, could explain the observed hybrid correlation, as argued by \cite{Coriat}.

\vspace{-1.em}
\subsection{The X-ray-loud hypothesis}
\label{sec:X-ray-bright}

An alternative approach is to invoke differences in the X-ray emission produced by the accretion flow between the two groups. It has been proposed that sources on the outlier track could be characterised by radiatively efficient accretion flows (e.g. \citealt{Coriat, Huang_2014}), such as the Luminous Hot Accretion Flow (LHAF, \citealt{Yuan_2004}) and would be brighter in X-rays than standard track sources. Those might in turn be modeled with the radiatively inefficient accretion flows, such as the Advection Dominated Accretion Flows (ADAF, e.g. \citealt{Narayan_1994}). Outliers could then be called \textit{X-ray-loud} instead of \textit{radio-quiet}.
Switching between tracks, hybrid sources would change from an an inefficient ADAF in the low flux regime to an efficient LHAF at high luminosity \citep{Coriat}. This is also supported
by the detection of hard X-ray cutoffs in the spectra of sources on the outlier track, implying
an effective cooling of the electrons responsible for the Comptonization of disk photons \citep{Koljonen}. It has been shown that the radiative efficiency of the accretion flow could depend on the mass accretion rate and hence on $L_{\tu{X}}$  \citep{Narayan_1995}, and a change from an ADAF to various types of LHAF is possible above a given critical luminosity  $L_{\tu{C}} \propto 5 \theta_{\tu{e}}^{3/2} \alpha_v^2 \dot{M}_{\tu{Edd}}$, where $\theta_{\tu{e}} = k_{\tu{b}} T_{\tu{e}}/m_{\tu{e}}c^2$ is the electron temperature in keV and $\alpha_v$ is the viscosity parameter of the accretion disk \citep{Xie_2012}. This would produce the outlier track. In the context of the \textit{accretion-jet} model \citep{Xie_2016}, different values of $\alpha_v$ lead to different $L_{\tu{C}}$, playing a key role in differentiating between standard track sources and outliers.
The path of \maxithirt{} agrees quite well with this scenario, since it is consistent with the standard track at $L_{\textup{X}} \lesssim 10^{33} (D/2.2 \ \tu{kpc})^2$ erg s$^{-1}$ and then moves to the flat part at the transition between the two tracks above $\sim$10$^{33} (D/2.2 \ \tu{kpc})^2$ erg s$^{-1}$, which could be a proxy for a low value of $L_{\textup{C}}$, and possibly a low $\alpha_v$. Future insights on the values of $\alpha_v$ (see for instance \citealt{Tetarenko_2018}) and $\theta_{\tu{e}}$ (e.g. \citealt{Koljonen}) would be crucial to quantify this agreement. A radiatively efficient accretion flow could also explain the behaviour of neutron star systems (NS XRBs), which are found to be generally fainter in radio than BH XRBs (e.g. \citealt{Migliari_2006, Tudor_2017}). At a given $L_{\textup{R}}$, the observed $L_{\textup{X}}$ might be increased by the additional X-ray emission coming from the solid surface. However, given the various nature of NS systems \citep{Tudor_2017, Gallo_2018}, the global situation appears to be more complex and a full comparison with NS XRBs is beyond the scope of this Letter. 

As an alternative, \cite{Meyer-Hofmeister_2014} suggested the presence of a weak, cool inner accretion disk, which would bring the source to the steep track by providing additional soft photons for Comptonization in the corona.
The inner disk, resulting from partial re-condensation of the coronal material, would be present in the hard state at $L_{\tu{X}} \gtrsim 10^{-4} L_{\tu{Edd}}$. Despite the difficulty of detecting such disk  with the current X-ray instruments, this scenario appears to reproduce well the path on the diagram followed by \hh{} \citep{Meyer-Hofmeister_2014}. However, \maxithirt{} leaves the standard track at $\sim$10$^{-6} (D/2.2 \ \tu{kpc})^2 L_{\tu{Edd}}$, and the existence of such disk is unlikely at these low accretion rates. 

While there is yet no consensus on what produces the observed tracks, it will be crucial for future monitorings of hybrid correlation sources to obtain a deep and dense coverage of the outburst, especially close to the transition luminosities and in the low-flux regime. Constraining the radio/X-ray correlation at low luminosities will allow us to discriminate between the existing models and to improve our understanding of the disk/jet connection among X-ray binaries.

\vspace{-1.em}
 
\section*{Acknowledgements}

We thank the anonymous referee for the careful reading of the manuscript and for providing valuable comments.
We thank the staff at the South African Radio Astronomy Observatory (SARAO) for scheduling these observations. The MeerKAT telescope is operated by the South African Radio Astronomy Observatory, which is a facility of the National Research Foundation, an agency of the Department of Science and Innovation. This work was carried out in part using facilities and data processing pipelines developed at the Inter-University Institute for Data Intensive Astronomy (IDIA). IDIA is a partnership of the Universities of Cape Town, of the Western Cape and of Pretoria. FC, SC and TDR thank Jamie Stevens and staff from the Australia Telescope National Facility (ATNF) for scheduling the ATCA radio observations. ATCA is part of the ATNF which is funded by the Australian Government for operation as a National Facility managed by CSIRO. We acknowledge the Gomeroi people as the traditional owners of the ATCA observatory site. We thank \emph{Swift} for scheduling the X-ray observations. We acknowledge the use of data obtained from the High Energy Astrophysics Science Archive Research Center (HEASARC), provided by NASA's Goddard Space Flight Center. This research has made use of MAXI data provided by RIKEN, JAXA and the MAXI team. FC acknowledges support from the project Initiative d’Excellence (IdEx) of Universit\'{e} de Paris (ANR-18-IDEX-0001). TDR acknowledges financial contribution from the agreement ASI-INAF n.2017-14-H.0.
We thank P. Saikia and W. Yu for their useful comments.
We acknowledge the use of the Nan\c cay Data Center, hosted by the Nan\c cay Radio Observatory (Observatoire de Paris-PSL, CNRS, Universit\'{e} d'Orl\'{e}ans), and supported by Region Centre-Val de Loire.

\vspace{-1.em}

\section*{Data availability}
The un-calibrated MeerKAT and ATCA visibility data are publicly available at the SARAO and ATNF archives, respectively at \url{https://archive.sarao.ac.za} and \url{https://atoa.atnf.csiro.au}. The \textit{Swift}/XRT data are instead available from the \textit{Swift} archive:
\url{https://www.swift.ac.uk/swift_portal}, while the MAXI data can be downloaded from \url{http://maxi.riken.jp/mxondem}.

\vspace{-1.em}

\bibliographystyle{mnras}

\bibliography{maxi1348_paper2}

\bsp	
\label{lastpage}
\end{document}